\documentclass[9pt,conference]{IEEEtran}
\IEEEoverridecommandlockouts

\usepackage{cite}
\usepackage{amsmath,amssymb,amsfonts}
\usepackage{graphicx}
\usepackage{bm}
\usepackage{booktabs} 
\usepackage{diagbox} 
\usepackage{enumitem}
\usepackage{multicol}
\usepackage{multirow}
\usepackage{makecell}
\usepackage{pifont}
\usepackage{subfigure}
\usepackage{xcolor}
\usepackage[linesnumbered,lined,ruled,commentsnumbered]{algorithm2e}
\usepackage{url,hyperref,cleveref}

\usepackage[normalem]{ulem}
\def\BibTeX{{\rm B\kern-.05em{\sc i\kern-.025em b}\kern-.08em
    T\kern-.1667em\lower.7ex\hbox{E}\kern-.125emXz}}

\begin{document}
\title{Streaming Keyword Spotting Boosted \\ by Cross-layer Discrimination Consistency}

\author{\IEEEauthorblockN{Yu Xi$^{1*}$, Haoyu Li$^{1*}$, Xiaoyu Gu$^2$, Hao Li$^2$, Yidi Jiang$^3$, Kai Yu$^{1{\dagger}}$ \thanks{$^{*}$ \text{Equal Contribution. $^{\dagger}$ Corresponding Author.} } }
\IEEEauthorblockA{\textit{$^1$MoE Key Lab of Artificial Intelligence, AI Institute, X-LANCE Lab, Shanghai Jiao Tong University, Shanghai, China}}
\IEEEauthorblockA{\textit{$^2$AISpeech Ltd, Suzhou, China}~~~~~\textit{$^3$National University of Singapore, Singapore}}
\IEEEauthorblockA{
\{yuxi.cs, haoyu.li.cs, kai.yu\}@sjtu.edu.cn~~~ \{xiaoyu.gu\_sx, hao.li\}@aispeech.com~~~ yidi\_jiang@u.nus.edu
}
}
\maketitle

\begin{abstract}

Connectionist Temporal Classification (CTC), a non-autoregressive training criterion, is widely used in online keyword spotting (KWS). However, existing CTC-based KWS decoding strategies either rely on Automatic Speech Recognition (ASR), which performs suboptimally due to its broad search over the acoustic space without keyword-specific optimization, or on KWS-specific decoding graphs, which are complex to implement and maintain. In this work, we propose a streaming decoding algorithm enhanced by Cross-layer Discrimination Consistency (CDC), tailored for CTC-based KWS. Specifically, we introduce a streamlined yet effective decoding algorithm capable of detecting the start of the keyword at any arbitrary position. Furthermore, we leverage discrimination consistency information across layers to better differentiate between positive and false alarm samples. Our experiments on both clean and noisy Hey Snips datasets show that the proposed streaming decoding strategy outperforms ASR-based and graph-based KWS baselines. The CDC-boosted decoding further improves performance, yielding an average absolute recall improvement of 6.8\% and a 46.3\% relative reduction in the miss rate compared to the graph-based KWS baseline, with a very low false alarm rate of 0.05 per hour.

\end{abstract}

\section{Introduction}
Keyword spotting (KWS), particularly for wake word detection (WWD), involves continuously detecting the presence of preset keywords in audio streams~\cite{icassp2014-guoguochen-dnn_kws,streaming-kws-2,streaming-kws-1,deep-spoken-kws-overview}. Given that KWS systems typically operate on resource-constrained devices, minimizing computational and memory costs is crucial. Connectionist Temporal Classification (CTC)~\cite{ctc}, which avoids the need for frame-level alignments, serves as an effective training criterion for this task. Numerous studies have explored various KWS applications within the CTC-based framework~\cite{is2016-yimengZhuang-ctc_kws_xlance_ctckws6,icassp2020-haikangYan-CRNN_ctc_kws_ctckws4,is2020-korea-kws_and_sv_ctckws5,icassp2021-yaotian-rnnt_fsmn_160khrs_kws_ByteDance_ctckws1,asru2023-aozhang-u2kws_ctckws2,arxiv2024-sichenJin-audio_text_embedding_openKWS_ctckws3}.

The non-autoregressive nature of CTC makes it inherently well-suited for designing efficient and streaming decoding algorithms. 
Existing decoding algorithms for continuous KWS can be categorized into two types. The first type is based on ASR inference~\cite{icassp2020-haikangYan-CRNN_ctc_kws_ctckws4,asru2023-aozhang-u2kws_ctckws2,taslp2021-runyanyang-ctc_las_keyword_search,icassp2023-jiewang-wekws}. In this approach, the model processes speech and outputs token sequences in a streaming mode, followed by matching the keywords within the updated decoding sequence. However, as noted in~\cite{icassp2024-yuxi-tdt_kws}, ASR decoding strategies~\cite{arxiv2014-ctc_prefix_search} aim to generate the most likely complete hypothesis across the entire space, rather than specifically detecting the presence of the keyword, making it suboptimal for KWS. The second type relies on the KWS decoding graph~\cite{hmm_filler_2_no_others,hmm_filler_4_no_others,arxiv2020-theodore-quantized_lstm_ctc_kws}, which typically includes both the keyword path and an optional filler path. While this method is tailored for KWS, implementing Weighted Finite State Transducer (WFST)-based graphs and Token Passing algorithms for decoding is complex, making them difficult to implement and enhance.

In~\cite{icassp2024-yuxi-tdt_kws}, the authors proposed an effective decoding algorithm for Transducer-based~\cite{RNNT} KWS, where they recursively fed decoded keyword token sequences rather than partial hypotheses into the Transducer predictor, restricting posterior generation and only decoding the keyword. Inspired by this approach, we propose a frame-synchronous streaming decoding algorithm specifically designed for CTC-based KWS. In addition, in our preliminary experiments, we observed significant differences in the behavior of intermediate and final layers around wake-up points for positive samples and easily triggered negative samples. To further enhance discrimination between challenging negative and positive samples, we leverage cross-layer discrimination consistency (CDC) information to propose the effective CDC-boosted streaming KWS decoding algorithm.

The key contributions of this paper are as follows:
\begin{enumerate}
    \item We propose a novel streaming decoding algorithm specifically designed for CTC-based KWS. Unlike ASR-based methods that search the entire space and can only initiate the search from the beginning of the utterance, our algorithm confines the search space to the keyword and allows detection at any time point. Additionally, it is simpler to implement compared to complex graph-based decoding methods.
    \item Based on the observation of distinct behavior across layers for positive and negative samples, we utilize cross-layer discrimination consistency to capture differences between keyword and non-keyword speech segments. Incorporating CDC further enhance the effectiveness of the CTC-KWS streaming decoding.
    \item The proposed CDC-boosted streaming KWS decoding algorithm shows impressive performance various SNR levels, compared to several ASR-based and graph-based baselines on the Hey Snips dataset. 
    
\end{enumerate}
\section{CDC-boosted streaming KWS}

\begin{figure}[t]
    \vspace{0em}
    \centerline{\includegraphics[width=1.0\columnwidth]{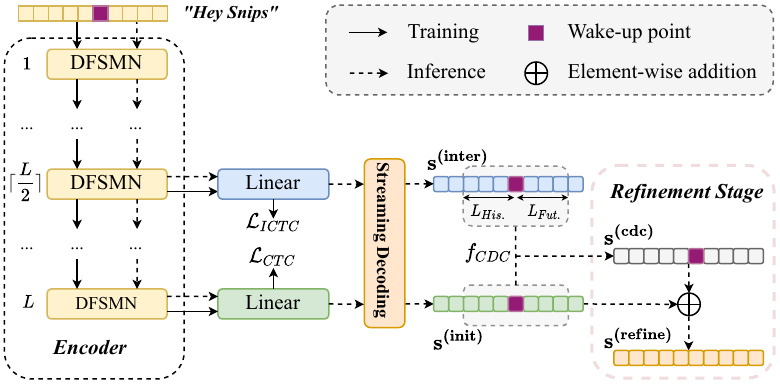}}
    \caption{An example of "Hey Snips" illustrates the training and decoding framework. $L$ represents the number of DFSMN layers. 
    The CDC-boosted decoding includes a refinement stage. 
    $\bm{s^{(init)}}$, $\bm{s^{(inter)}}$, $\bm{s^{(cdc)}}$, and $\bm{s^{(refine)}}$ denote the initial CTC decoding scores, ICTC decoding scores, CDC scores between initial and intermediate scores, and the final refined scores, respectively. $f_{CDC}$ represents the function to measure discrimination consitency. $L_{His.}$ and $L_{Fut.}$ refer to the history and future frame numbers used for CDC score computation.}
    \label{fig:overview}
\end{figure}

\subsection{CTC and Intermediate CTC}


Given the acoustic feature vectors $\mathbf{x} = [x_{1}, x_{2}, \cdots, x_{T}] \in \mathbb{R}^{T \times D}$ of an utterance, where $T$ is the number of frames and $D$ is the feature dimension, CTC models aim to predict the sequence of target labels $\mathbf{y} = [y_{1}, y_{2}, \cdots, y_{U}] \in \mathbb{R}^{U \times 1}$, where $U$ is the number of target labels. CTC accommodates repeated tokens and introduces a special blank token $\phi$ to facilitate alignment between the acoustic sequence and the label sequence. The alignment path is denoted as $\bm{\pi} = [\pi_1, \pi_2, \cdots, \pi_T] \in \mathbb{R}^{T \times 1}$. CTC seeks to maximize the likelihood of the target label sequence $\mathbf{y}$ over all possible alignment paths $\bm{\pi}$. The loss function can be formulated as:
\begin{align}
    \mathcal{L}_{CTC} &= -\log P(\mathbf{y} | \mathbf{x}) = -\log \sum_{\bm{\pi} \in \mathcal{B}^{-1}(\mathbf{y})} P(\bm{\pi}|\mathbf{x}).
\end{align}
Here, $\mathcal{B}$ represents the mapping from the alignment path $\bm{\pi}$ to the label sequence $\mathbf{y}$, while $\mathcal{B}^{-1}$ denotes the inverse mapping. All arithmetic operations are performed in the logarithmic domain.

Intermediate CTC~(ICTC) regularization~\cite{icassp2021-shinji-intermediate_ctc} is applied to the middle layer of the encoder as an effective regularization strategy~\cite{is2021-Jumon_Nozaki-ICTC_asr,icassp2022-yosuke-ICTC_asr,icassp2023-williamchen-ICTC_masr,icassp2024-maxime-avasr_ctcrnnt} for CTC training. As illustrated in~\Cref{fig:overview}, an additional linear projection layer with ICTC is attached to the middle section of the encoder. The overall loss function is formulated as follows:
\begin{align}
    \label{eq:overall-loss-function}
    \mathcal{L} = w \mathcal{L}_{ICTC} + (1-w) \mathcal{L}_{CTC},
\end{align}
where $w \in [0,1)$ is a hyper-parameter to tune the impact of ICTC.

\subsection{Streaming Decoding for CTC-based KWS}
\label{sec:2.2}

\begin{algorithm}[t]
    \caption{Streaming Decoding for CTC-based KWS}
    \label{algo:streaming-ctc}
    \KwIn{Posterior matrix $\mathbf{p} = [p_1,p_2,\cdots,p_T] \in \mathbb{R}^{T\times V}$, Keyword phoneme sequence $\mathbf{y}=[y_1,y_2,\cdots,y_U]$, Bonus score: $S_{bonus}$ and Timeout: $T_{out}$.} 
    \KwOut{$Score[T]$}  
    \BlankLine
    Init: 1) $\mathbf{\tilde{y}}=[\tilde{y}_{1},\tilde{y}_{2},\cdots,\tilde{y}_{\tilde{U}}]=[\phi,y_1,\phi,\cdots,\phi,y_{U},\phi]$ by inserting $\phi$ to $\mathbf{y}$, $\tilde{U} = len(\mathbf{\tilde{y}}) = 2U+ 1$
    

    
    Init: 2) $Score[1:T] = \{0\},\;\delta[1:T\,,1:\tilde{U}] = \{0\}$
    
    Init: 3) $\delta(1,1) = \delta(1,2) =1$
        
        

    
    \SetKw{Or}{or}
    \SetKw{Continue}{continue}
    
    \For{$t=2$ \Or $T$} {
        \If{ $\tilde{u} = 1 $ \Or $ \tilde{u} = 2 $ } {
        
        $\delta(t,\tilde{u})=1$ \# \textbf{new competitor starts at time $t$}
        
        \Continue
        }
        
        \For{$\tilde{u}=3$ \KwTo $\tilde{U}$} {
            \uIf { ($\tilde{y}_{\tilde{u}} == \phi$) } {
                $frd = \max\{\,\delta(t-1,\tilde{u}-1),\,\delta(t-1,\tilde{u})\,\}$
                
                $\delta(t,\tilde{u})= p_{t}(\phi) \cdot frd$
            }
            \Else {
                $frd = \max\{$
                
                $\quad \delta(t-1,\tilde{u}-2),\, \delta(t-1,\tilde{u}-1),\, \delta(t-1,\tilde{u})\,\}$
                
                
                $\delta(t,\tilde{u}) = p_{t}(\tilde{y}_{\tilde{u}}) \cdot frd$
            }
        }
        $Scores[t]=S_{Bonus} \cdot \max\{\,\delta(t,\tilde{U}-1),\, \delta(t,\tilde{U})\,\}$

        \BlankLine
    
        \# \textbf{record the length of the max score path as $\ell(t)$}

        $\ell(t) = RecordPathLength(Scores[t])$

        \BlankLine

    
        \If{ $\ell(t)> T_{out}$   } {  
            $Scores[t]=0$ \# \textbf{discard paths longer than $T_{out}$}
        } 

        $Scores[t]=pow(Scores[t],\, 1 / \ell(t))$ 
    }
    \Return $Scores[1:T]$ \# \textbf{return scores for each $t$}
\end{algorithm}

The core concept of KWS streaming decoding is to confine the search to the keyword at each time step. Unlike ASR decoding, which aims to generate a complete hypothesis, our method narrows the decoding space to the keyword-specific segment of the semantic space. 
Specifically, we target the phoneme posteriors related to the keyword and select the path with the highest likelihood of detecting the keyword or its partial occurrence at each decoding state. Traditional ASR or graph-based decoding methods are inflexible and require significant effort for online adaptation. In contrast, our search algorithm enables seamless keyword detection in streaming scenarios. At the start of each frame $t$, two new path competitors are introduced with probabilities initialized to 1, which enables seamless integration into the decoding process and ensures that new competitors entering at time step $t$ can effectively participate in the decoding. Given the potential length of the keyword, a timeout mechanism is also implemented to enhance the efficiency of the search. Further details are provided in~\Cref{algo:streaming-ctc}.

\subsection{Cross-layer Discrimination Consistency Boosted Decoding}

In addition to apply~\Cref{algo:streaming-ctc} on the main CTC layer, the shallow intermediate branch can also be leveraged for inference. We denote the initial and intermediate scores calculated by the algorithm proposed in \Cref{sec:2.2} from the main and intermediate CTC branches as:
\begin{align}
    \bm{s^{(init)}}=[\;s^{(init)}_1, s^{(init)}_2,\cdots,s^{(init)}_T\;]
    \label{eq:ctc-scores-merge},
\end{align}
and
\begin{align}
    \bm{s^{(inter)}}=[\;s^{(inter)}_1, s^{(inter)}_2,\cdots,s^{(inter)}_T\;]
    \label{eq:eq_decoding_3},
\end{align}
respectively.

\begin{figure}[t]
    \vspace{0em}
    \centerline{\includegraphics[width=1.0\columnwidth]{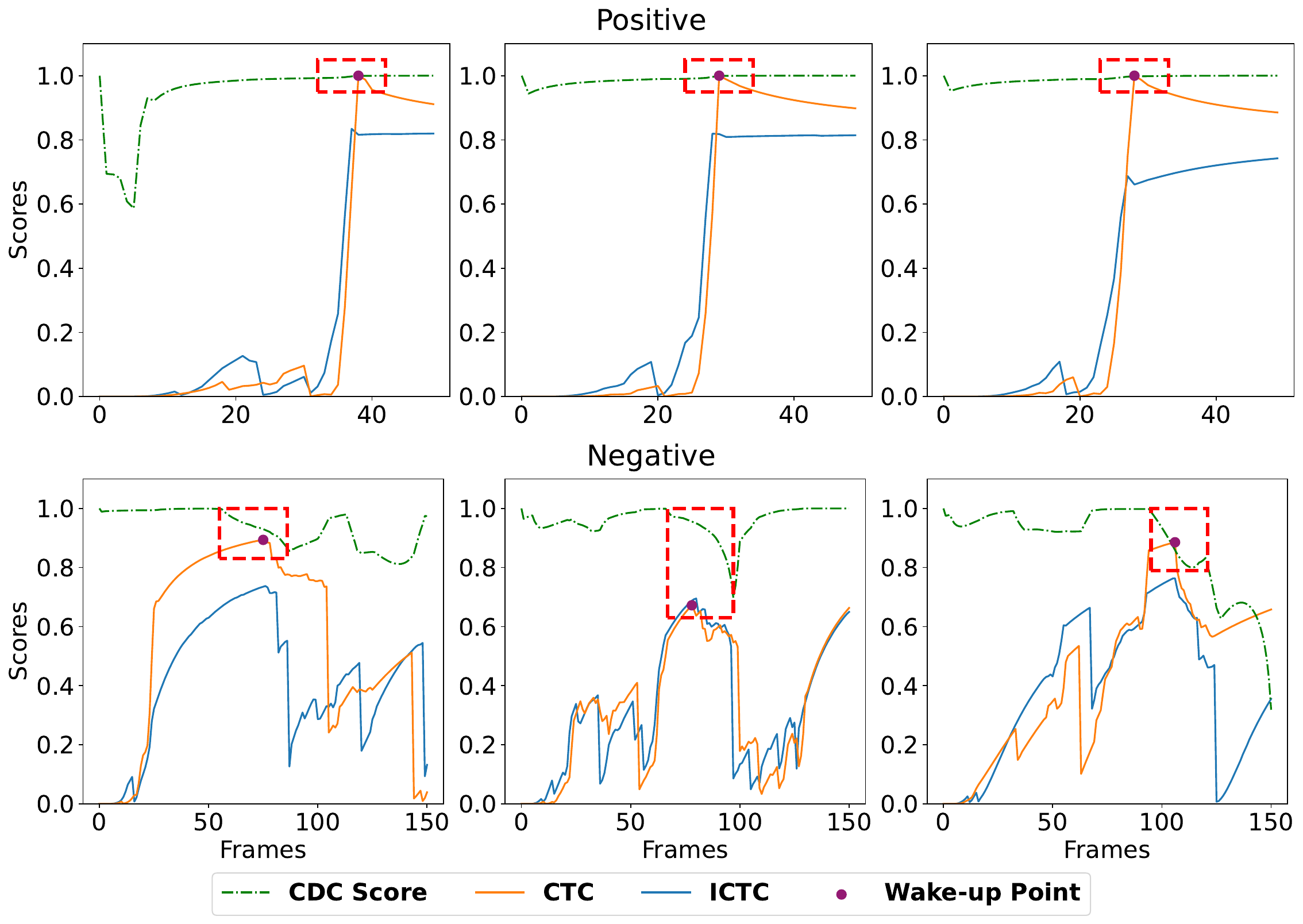} }
    \caption{The patterns of frame-level decoding scores are shown for positive (upper) and negative (lower) samples. Red boxes highlight changes in CDC scores near activation points. Notably, for positive samples, the scores consistently stay close to 1 around wake-up points, while negative samples show sharp variations.}
    \label{fig:kws-scores-patterns}
\end{figure}

Our preliminary experiments reveal that the scores from the main and intermediate layers exhibit distinct behaviors when the model processes positive versus negative data. As illustrated in~\Cref{fig:kws-scores-patterns}
, near the wake-up point, two score curves of positive data appear stable and similar. However, for high-score negative samples, the activation is brief, and the two score curves diverge significantly. This discrepancy likely arises because the patterns in positive samples are relatively consistent, allowing the ICTC layer to effectively learn these positive patterns, despite its limited parameters. However, negative samples encompass various patterns and conditions, making it challenging for shallow layers to fully capture all relevant aspects. Even if the final layer shows signs of overfitting to these false alarm patterns, the ICTC can still serve as a regularization mechanism. 

Then, considering the different behaviors for positive and negative samples, we define the \textit{sliding window CDC score} at time $t$ as the following:
\begin{equation}
\resizebox{0.91\hsize}{!}{
    $\bm{s^{(cdc)}_t} = f_{CDC} \left(\;\bm{s^{(init)}_{[(t-L_{His.}) : (t+L_{Fut.})]}},\; \bm{s^{(inter)}_{[(t-L_{His.}) : (t+L_{Fut.})]}\;} \right),$
}
\end{equation}
where $f_{CDC}$ denotes the function to measure scores of discrimination consistency, for which we employ the cosine similarity function here. $L_{His.}$ and $L_{Fut.}$ are history and future frame numbers. We also plot CDC score curves $\bm{s^{(cdc)}}$ in~\Cref{fig:kws-scores-patterns}. The refinement scores of the multi-stage decoding algorithm boosted by the CDC are the summation of initial CTC scores and the CDC scores: 
\begin{align}
    \bm{s^{(refine)}} = (\;\bm{s^{(init)}} + \bm{s^{(cdc)}}\;) \;/\; 2,
\end{align}
where $\bm{s^{(cdc)}}=[\,s^{(cdc)}_1, s^{(cdc)}_2,\cdots,s^{(cdc)}_T\,]$.
It is important to note that score calculation is performed frame-by-frame within a fully streaming decoding framework. The value of $L_{Fut.}$ can be adjusted to balance the trade-off between the performance of the refinement stage and system latency.

\section{Experimental Setups}
\subsection{Dataset}

Our experiments are conducted based on the following datasets.

\begin{itemize}
    \item \textbf{LibriSpeech}~\cite{LibriSpeech} is a large-scale English ASR dataset that contains approximately 960 hours of speech with corresponding transcription. We use LibriSpeech to initially train the CTC-based acoustic model to attain a superior foundation model.     
    \item \textbf{Hey-Snips (Snips)}~\cite{snips} is an open-source KWS dataset featuring the wake word ``Hey Snips." The positive portion of the dataset comprises 5,799, 2,484, and 2,599 keyword utterances in the train, dev, and test sets, respectively. Due to the absence of transcriptions, the negative samples cannot be utilized for ASR-based training. Consequently, we reorganize the Snips dataset as detailed in~\cite{icassp2024-yuxi-tdt_kws}. All original negative samples are compiled into a large negative test set, approximately 97 hours of speech.
    
    \item \textbf{WHAM!}~\cite{wham} is an ambient noise dataset that includes various kinds of ambient noise. We synthesize noisy keyword samples by clean Hey-Snips and noise randomly sampled from WHAM! to evaluate the robustness under different SNRs.
\end{itemize}

\subsection{Configuration}
\label{sec:feature}
\textbf{Training procedure.} We first pre-train our model using the full LibriSpeech dataset. During the fine-tuning phase, we sample an equal amount of data from LibriSpeech and Snips to construct a general ASR dataset. To enhance performance in noisy environments, each clean utterance from both Snips and LibriSpeech is mixed with a noise sample from WHAM!. The Signal-to-Noise Ratio (SNR) of these noisy utterances is uniformly distributed between 0 dB and 20 dB. In total, the fine-tuning dataset comprises approximately 24,000 utterances (12,000 clean and 12,000 noisy). 

\textbf{Acoustic features.} 
We extract 40-dimensional log Mel-filter bank coefficients (FBank) from each utterance using a window size of 25 ms and a hop size of 10 ms. Two types of data augmentation are applied during training: (1) online speech perturbation~\cite{Speed_Perturbation}, where a warping factor is uniformly sampled from $\{0.9, 1.0, 1.1\}$, and (2) SpecAugment~\cite{Specaug}, which uses 2 time-domain masks with $T_{max} = 50$ and 2 frequency-domain masks with $f_{max} = 10$. The FBank features from 5 preceding and 5 following frames are concatenated to create a 440-dimensional input. To reduce the input frame rate, a skipping factor of 3 frames is applied, resulting in each input frame capturing 30 ms of acoustic information.


\textbf{Model and loss.} The encoder comprises 6 DFSMN layers~\cite{DFSMN}, with input, hidden, and projection dimensions of 440, 512, and 320, respectively. The left-order and right-order of the DFSMN layers are set to 8 and 2. We utilize the CMU Pronouncing Dictionary ``cmudict-0.7b"~\cite{cmudict} to define the final acoustic output units. In addition to the 70 monophones, a unique blank token is included for CTC modeling. During the pre-training phase, we use the frame-level cross-entropy criterion exclusively. In the fine-tuning phase, either CTC alone or CTC with ICTC is applied, depending on the model. Unless otherwise specified, the weight $w$ in~\Cref{eq:overall-loss-function} for ICTC is set to 0.3.

\subsection{Evaluation} 

\textbf{Evaluation procedure.} 
To accurately assess performance across different SNR scenarios, each positive sample is mixed with noise at SNR levels of $\{0, 5, 10, 15, 20\}$ dB. Additionally, a positive set with -5 dB is also created to evaluate the system's robustness under extremely low and out-of-domain SNR conditions. As previously mentioned, all negative samples from the original Snips dataset are gathered for testing. The negative dataset is also mixed with noise, with SNRs ranging from 0 dB to 20 dB. Decoding hyper-parameters $S_{Bonus}$ and $T_{out}$ in~\Cref{algo:streaming-ctc} are set to $e^3$ and 3 seconds, respectively.


\textbf{Evaluation metrics.} We evaluate recall (1-miss rate) and macro-averaged recall across different SNR levels at a fixed false alarm rate (FAR). For ASR-based methods, we report accuracy (Acc.), which corresponds to recall at FAR=0, due to the challenges in decoding false alarms for these systems. In addition, when comparing with KWS graph-based systems, we present results at a FAR of 0.05 per hour, a more reasonable yet stringent metric for KWS systems.

\section{Results and Analysis}

\subsection{Streaming Decoding for KWS}

In~\Cref{tab:asr-vs-kws}, we compare three mainstream decoding baselines with the same CTC-based acoustic model. The CTC greedy decoding and prefix beam search~\cite{arxiv2014-ctc_prefix_search,icassp2023-jiewang-wekws} are both ASR-based systems. 
Our streaming method significantly outperforms all three ASR-based and graph-based baselines. Compared to the two ASR-based baselines, streaming KWS decoding achieves absolute improvements of 10.4\% and 6.6\% across different SNR levels. Even against the powerful graph-based baseline, our streaming decoding strategy delivers an absolute gain of 6.3\%. The results indicate that the proposed method can perform well across different noisy environments.

There are several important details to consider. The proposed method achieves an absolute gain of around or over 10\% in accuracy under low SNR conditions, such as -5 or 0 dB. This improvement may be because, at very low SNRs, ASR-based decoding struggles to generate correct hypotheses due to excessive noise in the speech. For graph-based methods, decoding tokens find it difficult to pass through the decoding graph, which prevents them from reaching final states and results in low accuracy for graph-based baselines. However, our method enforces decoding to output activation scores frame-by-frame, allowing it to continue despite strong background noise. This demonstrates that our approach is particularly effective in challenging acoustic environments.

d
\begin{table}[t]
    \centering
    \caption{Accuracy comparison (FAR=0) of the proposed streaming algorithm and 3 baseline systems:  (1) ASR greedy search, (2) ASR prefix beam search and (3) KWS graph-based search.}
    \begin{resizebox}{1.0\columnwidth}{!}{
        \begin{tabular}{c|c|cccccc|c}
            \toprule
             \multirow{2}{*}{\textbf{Decoding Alg.}} & \multicolumn{7}{c|}{\textbf{SNR}} & \multirow{2}{*}{\textbf{Avg.}} \\
            \cmidrule(lr){2-8}
            & -5 & 0 & 5 & 10 & 15 & 20 & +inf & \\            
            \midrule
            Greedy Search & 22.7 & 50.7 & 73.1 & 84.1 & 88.7 & 90.5 & 91.9 & 71.7  \\
            Prefix Beam Search (beam = 10) & 29.1 & 58.5 & 78.1 & 87.1 & 90.4 & 91.9 & 93.1 & 75.5 \\
            KWS Graph & 23.9 & 54.7 & 79.1 & 88.7 & 93.1 & 94.7 & 96.2 & 75.8 \\
            \midrule
            Streaming KWS & \textbf{35.5} & \textbf{68.0} & \textbf{86.2} & \textbf{93.6} & \textbf{96.3} & \textbf{97.4} & \textbf{97.7} & \textbf{82.1} \\
            \bottomrule
        \end{tabular}
    }\end{resizebox}
    \label{tab:asr-vs-kws}
\end{table}


\subsection{CDC-boosted Streaming Decoding}


This section highlights the importance of the CDC strategy for improving the streaming decoding algorithm. As shown in~\Cref{tab:main-table}, when comparing pairs that use the same decoding strategy, such as (A vs. C) and (B vs. D), we observe that the gains from ICTC are more pronounced at low SNR levels, indicating the effectiveness of CTC regularization for noisy scenarios. Additionally, the comparison between models (E vs. D) further demonstrates the effectiveness of our proposed CDC strategy.
Overall, when comparing the CDC boosted system (E) with the baseline (A), we achieve an absolute 6.8\% increase in average recall and a 46.3\% relative reduction in average miss rate across various SNR levels.

\begin{table}[h!]
    \centering
    \caption{The recall comparison of the proposed KWS streaming and multi-stage decoding strategies with graph-based baselines across various in-domain and out-of-domain SNR levels at a fixed FAR of 0.05 per hour.}
    \begin{resizebox}{1.0\columnwidth}{!}{
        
    \begin{tabular}{c|cc|c|cccccc|c}
        \toprule

        \multirow{2}{*}{\textbf{ID}} & \multirow{2}{*}{{$\bm{\mathcal{L}_{ICTC}}$}} & \multirow{2}{*}{ \textbf{Decoding Alg.} }  & \multicolumn{7}{c|}{\textbf{SNR}} & \multirow{2}{*}{\textbf{Avg.}} \\
        \cmidrule(lr){4-10}
         & &  & -5 & 0 & 5 & 10 & 15 & 20 & +inf &  \\            
        \midrule
        A & \multirow{2}{*}{\ding{56}} & KWS Graph & 41.6 & 75.1 & 90.2 & 95.8 & 97.5 & 98.3 & 98.8 & 85.3 \\ 
        B &  & Streaming & 54.0 & 82.7 & 93.9 & 97.6 & 98.9 & 99.3 & 99.4 & 89.4 \\
        \midrule
        
       C &  \multirow{3}{*}{\ding{51}} & KWS Graph & 44.4 & 76.7 & 90.2 & 95.3 & 96.8 & 97.5 & 98.3 & 85.6 \\
       D &  & Streaming & 59.8 & 86.4 & 95.4 & 98.2 & 98.9 & 99.2 & 99.7 & 91.1 \\

        E & & CDC-boosted Streaming  & \textbf{63.4} & \textbf{88.4} & \textbf{95.9} & \textbf{98.7} & \textbf{99.2} & \textbf{99.5} & \textbf{99.8} & \textbf{92.1} \\

        \bottomrule
    \end{tabular}
    }\end{resizebox}

    \label{tab:main-table}
\end{table}

\subsection{Analysis on History/Future Sizes of CDC}

The CDC strategy leverages score patterns around the wake-up point, which inevitably introduces some latency.
Therefore, we further analyze the impact of the history window size ($L_{His.}$) and the future window size ($L_{Fut.}$) on KWS performance. As mentioned in~\Cref{sec:feature}, each frame contains 30 ms of acoustic information. We fix the total window size to 30 frames, resulting in a maximum latency of 900 ms, which is acceptable for deployment.
In~\Cref{tab:window-size-and-performance}, it is unsurprising that increasing the number of future frames improves performance, highlighting the effectiveness of discrimination information provided by future features around activation points. Additionally, we present two upper-bound baselines: one uses all future scores, and another is a completely offline baseline using all scores from the entire utterance. The results show that the performance gain beyond $L_{Fut.}=30$ is minimal, indicating that excessive discrimination information offers diminishing returns. Introducing a slight latency (less than 1 second) is sufficient to achieve near upper-bound performance gains.


\begin{table}[t]
    \centering
    \caption{The comparison of recall performance using different windows in the CDC-boosted decoding. We bold and underline \uline{\textbf{the results of the proposed system}}, and bold \textbf{the upper-bound offline results} for reference.}
    \resizebox{1.0\columnwidth}{!}{
    \begin{tabular}{cc|c|c|cccccc|c}
        \toprule
        \multirow{2}{*}{$\bm{L_{His.}}$} & \multirow{2}{*}{$\bm{L_{Fut.}}$} & \multirow{2}{*}{\makecell{ \textbf{Lat.} \\ \textbf{(ms)} }} & \multicolumn{7}{c|}{\textbf{SNR}} & \multirow{2}{*}{\textbf{Avg.}} \\
        \cmidrule(lr){4-10}
         & & & -5 & 0 & 5 & 10 & 15 & 20 & +inf & \\
        \midrule
        0 & 30 & 900 & \uline{\textbf{63.4}} & \uline{\textbf{88.4}} & \uline{\textbf{95.9}} & \uline{\textbf{98.7}} & \uline{\textbf{99.2}} & \uline{\textbf{99.5}} & \uline{\textbf{99.8}} & \uline{\textbf{92.1}} \\
                \midrule
        5 & 25 & 750 & 61.0 & 87.3 & 95.5 & 98.5 & 99.1 & 99.4 & 99.7 & 91.5 \\
        10 & 20 & 600 & 60.4 & 86.9 & 95.4 & 98.3 & 99.0 & 99.3 & 99.7 & 91.3 \\
        20 & 10 & 300 & 59.6 & 86.1 & 95.2 & 98.2 & 99.0 & 99.3 & 99.6 & 91.0 \\
        30 & 0 & 0 & 58.4 & 85.6 & 95.0 & 98.1 & 98.9 & 99.2 & 99.6 & 90.7 \\

        \midrule
        0 & $\infty$ & $\infty$ & 64.0 & \textbf{88.3} & 95.8 & 98.5 & 99.2 & 99.5 & 99.7 & 92.1 \\
        $\infty$ & $\infty$ & $\infty$ & \textbf{64.3} & 88.2 & \textbf{96.1} & \textbf{98.7} & \textbf{99.3} & \textbf{99.6} & \textbf{99.8} & \textbf{92.3} \\
        \bottomrule
    \end{tabular}
    }
    \label{tab:window-size-and-performance}
\end{table}


\subsection{Optimal Intermediate Regularization Setup}

We analyze the impact of the intermediate loss weight $w$ in~\Cref{eq:overall-loss-function} and the layer to which ICTC is applied. The results in~\Cref{tab:w-and-applied-layer} show that, with a fixed layer, a value of $w = 0.3$ optimally balances the contribution of $\mathcal{L}_{ICTC}$. Additionally, experiments on the applied layer suggest that placing ICTC around $L/2$ achieves the best performance. This is because too shallow layers focus primarily on feature extraction and do not effectively capture keyword information, while layers that are too deep learn information similar to the main CTC. Therefore, neither is ideal for extracting CDC information.

\begin{table}[h!]
    \centering
    \caption{The recall results for different $w$ and the layer applied ICTC. The encoder consists of a total of $L = 6$ layers, with ${L}/{2} = 3$.}
    \label{tab:w-and-applied-layer}
    \resizebox{1.0\columnwidth}{!}{
    \begin{tabular}{c|c|c|cccccc|c}
         \toprule
         \multirow{2}{*}{\makecell{\textbf{Layer}\\ \textbf{ID}}} & \multirow{2}{*}{{$\bm{w}$}} & \multicolumn{7}{c|}{\textbf{SNR}} & \multirow{2}{*}{\textbf{Avg.}} \\
        \cmidrule(lr){3-9}
          & & -5 & 0 & 5 & 10 & 15 & 20 & +inf & \\
         \midrule
         3 & 0.3 & 63.4 & 88.4 & 95.9 & 98.7 & 99.2 & 99.5 & 99.8 & 92.1 \\
        \midrule
         3 & 0.4 & 61.9 & 87.7 & 95.8 & 98.6 & 99.2 & 99.4 & 99.7 & 91.8 \\
         3 & 0.2 & 60.4 & 86.9 & 95.7 & 98.2 & 99.2 & 99.6 & 99.8 & 91.4 \\
         3 & 0.1 & 59.8 & 86.1 & 95.6 & 98.4 & 99.3 & 99.6 & 99.9 & 91.2 \\
         \midrule
         2 & 0.3 & 53.3 & 82.0 & 93.8 & 97.9 & 98.9 & 99.4 & 99.7 & 89.3 \\
         4 & 0.3 & 64.0 & 88.7 & 96.1 & 98.7 & 99.3 & 99.6 & 99.7 & 92.3 \\
         5 & 0.3 & 59.8 & 86.4 & 95.5 & 98.4 & 99.2 & 99.3 & 99.6 & 91.2 \\
         \bottomrule
    \end{tabular}
    }
\end{table}

\section{Conclusion}
In this paper, we propose the CDC-boosted streaming KWS decoding algorithm designed for CTC-based systems. Specifically, we propose a streaming decoding algorithm which can detect arbitrary start point of the keyword in audio stream and further boost the performance by leveraging CDC scores derived from the intermediate and final layer. Our streaming decoding approach achieves absolute improvements of 10.4\% and 6.6\% in accuracy compared to the greedy and prefix beam search CTC ASR baselines. Furthermore, the CDC-boosted version provides an absolute gain of 6.8\% in recall and a 46.3\% relative reduction in miss rate at a very low FAR of 0.05 per hour, outperforming the SOTA graph-based CTC KWS method. Our method not only delivers superior performance but also demonstrates robust performance under extremely low and out-of-domain SNR conditions. Moreover, it is easy to implement and maintain, making it well-suited for practical applications and further improvements.

\clearpage
\bibliographystyle{style/IEEEtran}
\bibliography{citations/refs}

\begin{thebibliography}{10}
\providecommand{\url}[1]{#1}
\csname url@samestyle\endcsname
\providecommand{\newblock}{\relax}
\providecommand{\bibinfo}[2]{#2}
\providecommand{\BIBentrySTDinterwordspacing}{\spaceskip=0pt\relax}
\providecommand{\BIBentryALTinterwordstretchfactor}{4}
\providecommand{\BIBentryALTinterwordspacing}{\spaceskip=\fontdimen2\font plus
\BIBentryALTinterwordstretchfactor\fontdimen3\font minus \fontdimen4\font\relax}
\providecommand{\BIBforeignlanguage}[2]{{%
\expandafter\ifx\csname l@#1\endcsname\relax
\typeout{** WARNING: IEEEtran.bst: No hyphenation pattern has been}%
\typeout{** loaded for the language `#1'. Using the pattern for}%
\typeout{** the default language instead.}%
\else
\language=\csname l@#1\endcsname
\fi
#2}}
\providecommand{\BIBdecl}{\relax}
\BIBdecl

\bibitem{icassp2014-guoguochen-dnn_kws}
G.~Chen, C.~Parada, and G.~Heigold, ``Small-footprint keyword spotting using deep neural networks,'' in \emph{Proc. IEEE ICASSP}, 2014, pp. 4087--4091.

\bibitem{streaming-kws-2}
R.~Alvarez and H.-J. Park, ``End-to-end streaming keyword spotting,'' in \emph{Proc. IEEE ICASSP}, 2019, pp. 6336--6340.

\bibitem{streaming-kws-1}
Y.~Wang, H.~Lv, D.~Povey, L.~Xie, and S.~Khudanpur, ``Wake word detection with streaming transformers,'' in \emph{Proc. IEEE ICASSP}, 2021, pp. 5864--5868.

\bibitem{deep-spoken-kws-overview}
I.~López-Espejo, Z.-H. Tan, J.~H.~L. Hansen, and J.~Jensen, ``Deep spoken keyword spotting: An overview,'' \emph{IEEE Access}, pp. 4169--4199, 2022.

\bibitem{ctc}
A.~Graves, S.~Fern{\'{a}}ndez, F.~J. Gomez, and J.~Schmidhuber, ``Connectionist temporal classification: labelling unsegmented sequence data with recurrent neural networks,'' in \emph{Proc. ICML}, 2006, pp. 369--376.

\bibitem{is2016-yimengZhuang-ctc_kws_xlance_ctckws6}
Y.~Zhuang, X.~Chang, Y.~Qian, and K.~Yu, ``Unrestricted vocabulary keyword spotting using lstm-ctc.'' in \emph{Proc. ISCA Interspeech}, 2016, pp. 938--942.

\bibitem{icassp2020-haikangYan-CRNN_ctc_kws_ctckws4}
H.~Yan, Q.~He, and W.~Xie, ``{CRNN}-{CTC} based mandarin keywords spotting,'' in \emph{Proc. IEEE ICASSP}, 2020, pp. 7489--7493.

\bibitem{is2020-korea-kws_and_sv_ctckws5}
M.~Jung, Y.~Jung, J.~Goo, and H.~Kim, ``Multi-task network for noise-robust keyword spotting and speaker verification using {CTC}-based soft {VAD} and global query attention,'' in \emph{Proc. ISCA Interspeech}, 2020, pp. 931--935.

\bibitem{icassp2021-yaotian-rnnt_fsmn_160khrs_kws_ByteDance_ctckws1}
Y.~Tian, H.~Yao, M.~Cai, Y.~Liu, and Z.~Ma, ``Improving rnn transducer modeling for small-footprint keyword spotting,'' in \emph{Proc. IEEE ICASSP}, 2021, pp. 5624--5628.

\bibitem{asru2023-aozhang-u2kws_ctckws2}
A.~Zhang, P.~Zhou, K.~Huang, Y.~Zou, M.~Liu, and L.~Xie, ``U2-{KWS}: Unified two-pass open-vocabulary keyword spotting with keyword bias,'' in \emph{Proc. IEEE ASRU}, 2023, pp. 1--8.

\bibitem{arxiv2024-sichenJin-audio_text_embedding_openKWS_ctckws3}
S.~Jin, Y.~Jung, S.~Lee, J.~Roh, C.~Han, and H.~Cho, ``{CTC}-aligned audio-text embedding for streaming open-vocabulary keyword spotting,'' \emph{CoRR}, 2024.

\bibitem{taslp2021-runyanyang-ctc_las_keyword_search}
R.~Yang, G.~Cheng, H.~Miao, T.~Li, P.~Zhang, and Y.~Yan, ``Keyword search using attention-based end-to-end asr and frame-synchronous phoneme alignments,'' \emph{IEEE/ACM Transactions on Audio, Speech, and Language Processing}, vol.~29, pp. 3202--3215, 2021.

\bibitem{icassp2023-jiewang-wekws}
J.~Wang, M.~Xu, J.~Hou, B.~Zhang, X.~Zhang, L.~Xie, and F.~Pan, ``We{KWS}: {A} production first small-footprint end-to-end keyword spotting toolkit,'' in \emph{Proc. IEEE ICASSP}, 2023, pp. 1--5.

\bibitem{icassp2024-yuxi-tdt_kws}
Y.~Xi, H.~Li, B.~Yang, H.~Li, H.~Xu, and K.~Yu, ``{TDT-KWS}: Fast and accurate keyword spotting using token-and-duration transducer,'' in \emph{Proc. IEEE ICASSP}, 2024, pp. 11\,351--11\,355.

\bibitem{arxiv2014-ctc_prefix_search}
\BIBentryALTinterwordspacing
A.~L. Maas, A.~Y. Hannun, D.~Jurafsky, and A.~Y. Ng, ``First-pass large vocabulary continuous speech recognition using bi-directional recurrent dnns,'' \emph{CoRR}, vol. abs/1408.2873, 2014. [Online]. Available: \url{http://arxiv.org/abs/1408.2873}
\BIBentrySTDinterwordspacing

\bibitem{hmm_filler_2_no_others}
M.~Sun, D.~Snyder, Y.~Gao, V.~K. Nagaraja, M.~Rodehorst, S.~Panchapagesan, N.~Strom, S.~Matsoukas, and S.~Vitaladevuni, ``Compressed time delay neural network for small-footprint keyword spotting.'' in \emph{Proc. Interspeech}, 2017, pp. 3607--3611.

\bibitem{hmm_filler_4_no_others}
M.~Wu, S.~Panchapagesan, M.~Sun, J.~Gu, R.~Thomas, S.~N.~P. Vitaladevuni, B.~Hoffmeister, and A.~Mandal, ``Monophone-based background modeling for two-stage on-device wake word detection,'' in \emph{Proc. IEEE ICASSP}, 2018, pp. 5494--5498.

\bibitem{arxiv2020-theodore-quantized_lstm_ctc_kws}
\BIBentryALTinterwordspacing
T.~Bluche, M.~Primet, and T.~Gisselbrecht, ``Small-footprint open-vocabulary keyword spotting with quantized {LSTM} networks,'' \emph{CoRR}, vol. abs/2002.10851, 2020. [Online]. Available: \url{https://arxiv.org/abs/2002.10851}
\BIBentrySTDinterwordspacing

\bibitem{RNNT}
A.~Graves, ``Sequence transduction with recurrent neural networks,'' in \emph{ICML — Workshop on Representation Learning}, 2012.

\bibitem{icassp2021-shinji-intermediate_ctc}
J.~Lee and S.~Watanabe, ``Intermediate loss regularization for {CTC}-based speech recognition,'' in \emph{Proc. IEEE ICASSP}, 2021, pp. 6224--6228.

\bibitem{is2021-Jumon_Nozaki-ICTC_asr}
J.~Nozaki and T.~Komatsu, ``Relaxing the conditional independence assumption of ctc-based {ASR} by conditioning on intermediate predictions,'' in \emph{Proc. ISCA Interspeech}, 2021, pp. 3735--3739.

\bibitem{icassp2022-yosuke-ICTC_asr}
Y.~Higuchi, K.~Karube, T.~Ogawa, and T.~Kobayashi, ``Hierarchical conditional end-to-end {ASR} with {CTC} and multi-granular subword units,'' in \emph{Proc. IEEE ICASSP}, 2022, pp. 7797--7801.

\bibitem{icassp2023-williamchen-ICTC_masr}
W.~Chen, B.~Yan, J.~Shi, Y.~Peng, S.~Maiti, and S.~Watanabe, ``Improving massively multilingual {ASR} with auxiliary {CTC} objectives,'' in \emph{Proc. IEEE ICASSP}, 2023, pp. 1--5.

\bibitem{icassp2024-maxime-avasr_ctcrnnt}
M.~Burchi, K.~C. Puvvada, J.~Balam, B.~Ginsburg, and R.~Timofte, ``Multilingual audio-visual speech recognition with hybrid {CTC/RNN-T} fast conformer,'' in \emph{Proc. IEEE ICASSP}.\hskip 1em plus 0.5em minus 0.4em\relax {IEEE}, 2024, pp. 10\,211--10\,215.

\bibitem{LibriSpeech}
V.~Panayotov \emph{et~al.}, ``Librispeech: an asr corpus based on public domain audio books,'' in \emph{Proc. IEEE ICASSP}, 2015, pp. 5206--5210.

\bibitem{snips}
A.~Coucke, M.~Chlieh, T.~Gisselbrecht, D.~Leroy, M.~Poumeyrol, and T.~Lavril, ``Efficient keyword spotting using dilated convolutions and gating,'' in \emph{Proc. IEEE ICASSP}, 2019, pp. 6351--6355.

\bibitem{wham}
G.~Wichern, J.~Antognini, M.~Flynn, L.~R. Zhu, E.~McQuinn, D.~Crow, E.~Manilow, and J.~L. Roux, ``Wham!: Extending speech separation to noisy environments,'' in \emph{Proc. ISCA Interspeech}, 2019, pp. 1368--1372.

\bibitem{Speed_Perturbation}
T.~Ko, V.~Peddinti, D.~Povey, and S.~Khudanpur, ``Audio augmentation for speech recognition,'' in \emph{Proc. ISCA Interspeech}, 2015.

\bibitem{Specaug}
D.~S. Park, W.~Chan, Y.~Zhang, C.-C. Chiu, B.~Zoph, E.~D. Cubuk, and Q.~V. Le, ``Specaugment: A simple data augmentation method for automatic speech recognition,'' in \emph{Proc. ISCA Interspeech}, 2019.

\bibitem{DFSMN}
S.~Zhang \emph{et~al.}, ``Deep-{FSMN} for large vocabulary continuous speech recognition,'' in \emph{Proc. IEEE ICASSP}.\hskip 1em plus 0.5em minus 0.4em\relax {IEEE}, 2018, pp. 5869--5873.

\bibitem{cmudict}
``The {CMU} pronouncing dictionary,'' \url{http://www.speech.cs.cmu.edu/cgi-bin/cmudict}.

\end{thebibliography}
\end{document}